\documentclass[aps,preprint,nofootinbib]{revtex4}
\usepackage{amsfonts}
\usepackage{graphicx,calc,epsfig}

\newcommand{\C}{\mathbb{C}}
\newcommand{\R}{\mathbb{R}}
\newcommand{\be}{\nopagebreak[3]\begin{equation}}
\newcommand{\ee}{\end{equation}}
\newcommand{\ba}{\nopagebreak[3]\begin{eqnarray}}
\newcommand{\ea}{\end{eqnarray}}
\DeclareFontFamily{U}{rsfs}{}         
\DeclareFontShape{U}{rsfs}{m}{n}{<5> rsfs5 <6><7> rsfs7          %
  <8><9><10><10.95><12><14.4><17.28><20.74><24.88> rsfs10}{}     %
\DeclareMathAlphabet{\mathfs}{U}{rsfs}{m}{n}                     %
\newcommand{\mfs}[1]{\mathfs {#1}}                               %
\newcommand{\va}{\scriptscriptstyle}

\newcommand{\sH}{{\mfs H}}

\newcommand{\Ss}{{\mfs S}}
\newcommand{\sM}{{\mfs M}}
\newcommand{\sB}{{\mfs B}}
\newcommand{\sW}{{\mfs W}}
\newcommand{\Hp}{{\sH}_{phys}}

\newcommand{\Hk}{{\sH}_{kin}}

\newcommand{\Ha}{{\sH}_{aux}}
\newcommand{\Hb}{{\sH}}

\newcommand{\g}{\mathfrak{g}}
\newcommand{\su}{\mathfrak{su}}
\newcommand{\SU}{\mathrm{SU}}
\newcommand{\SO}{\mathrm{SO}}

\newcommand{\tr}{\mathrm{tr}}

\newcommand{\Irrep}{\mathrm{Irrep}}
\newcommand{\Cyl}{{Cyl}}

\begin{document}

\title{Quantization of strings and branes coupled to $BF$ theory}
\author{John C.\ Baez}
\affiliation{Department of Mathematics,\\ University of
California,\\ Riverside, CA 92521, USA. }
\author{Alejandro Perez}
\affiliation{Centre de Physique Th\'eorique, Unit\'e Mixte de
Recherche (UMR 6207) du CNRS et des Universit\'es Aix-Marseille I,
Aix-Marseille II, et du Sud Toulon-Var; laboratoire afili\'e \`a
la FRUMAM (FR 2291), Campus de Luminy, 13288 Marseille,
France.}

\date{\today \vbox{\vskip 2em}}

\begin{abstract}
$BF$ theory is a topological theory that can be seen as a natural
generalization of 3-dimensional gravity to arbitrary
dimensions.  Here we show that the coupling to point particles that
is natural in three dimensions generalizes in a direct way to $BF$
theory in $d$ dimensions coupled to $(d-3)$-branes.
In the resulting model, the connection is flat except along the membrane
world-sheet, where it has a conical singularity whose strength is
proportional to the membrane tension.  As a step towards canonically
quantizing these models, we show that a basis of kinematical states
is given by `membrane spin networks', which are spin networks
equipped with extra data where their edges end on a brane.
\end{abstract}

\maketitle

Interest in the quantization of $2+1$ gravity coupled to point
particles has been revived in the context of the spin foam
\cite{spinfoam} and loop quantum gravity \cite{lqg} approaches
to the nonperturbative and background-independent
quantization of gravity.  On the one hand this simple
system provides a nontrivial example where the strict relation
between the covariant and canonical approaches can be
demonstrated \cite{ky}. On the other hand intriguing relationships
with field theories with infinitely many degrees of freedom have
been obtained \cite{lau}.

The idea of generalizing this
construction to higher dimensions is very appealing.
We will argue that in $3+1$ dimensions, the natural objects
replacing point particles are strings.  This
idea has already been studied in a companion paper \cite{baez},
which treated these strings merely as defects in the gauge field---
i.e., places where it has a conical singularity.
Here we propose a specific dynamics for the theory and a strategy
for quantizing it.  More generally, in $d$-dimensional spacetime we
describe a way to couple $(d-3)$-branes to $BF$ theory.

To understand this, first recall that in three dimensions,
Einstein's equations force the curvature to vanish at every point
of spacetime.  Therefore, except for global topological
excitations, three-dimensional pure gravity does not have local
degrees of freedom.  However, it is precisely this local rigidity
of Einstein's gravity in three dimensions that makes it easy to
couple the theory to point particles.  The presence of massive
point particles in three-dimensional gravity modifies the
classical solutions by producing conical curvature singularities
along the particles' world-lines.  With this idea in mind, one can
write an action for a single particle coupled to gravity by
introducing a source term to the standard action in the first
order formalism, namely: \be \label{naive3d} S(A,e)=\int_{\sM}
\tr[e\wedge F(A)])+ m \int_{\gamma}\tr[e\,v], \ee where $m$ is the
mass of the particle, $v$ is a fixed unit vector in the Lie
algebra $\su(2)$, and $\gamma$ is the world-line of the particle.
It is easy to see that the previous action leads to equations of
motion whose solutions are flat everywhere except for a conical
singularity along $\gamma$, as desired.

Unfortunately, this action suffers two drawbacks. First, it is no
longer invariant under the standard gauge symmetries of pure
gravity.  Second, there is no explicit dependence on the particle
degrees of freedom: one is describing the particle simply as a
gauge defect along $\gamma$. One can solve both problems in one
stroke by adding degrees of freedom for the particles, and
choosing an action invariant under an appropriate extension of the
gauge group of the system. The result is the Sousa Gerbert action
\cite{1} for a spinless point particle of mass $m$ coupled to
three-dimensional Riemannian gravity: 
\be 
\label{better3d}
S(A,e,q,\lambda)=\int_{\sM} \tr[ e\wedge F(A)] + m \int_{\gamma}
\tr[(e + d_A q) \, \lambda v\lambda^{-1}]. 
\ee 
Here $v$ is a fixed
unit vector in $\su(2)$ as before, while the particle's degrees of
freedom are described by an $\su(2)$-valued function $q$ and an
$\SU(2)$-valued function $\lambda$ defined on the world-line
$\gamma$. The physical interpretation of $q$ is a bit obscure, but
we can think of it as `position in an internal space'.    In a
similar way, $p = m \lambda v \lambda^{-1}$ represents the particle's
momentum, which is treated as an independent variable in this
first-order formulation.

This action is invariant under the gauge transformations
\be
\label{gravy1} 
\begin{array}{ccl}
e &\mapsto & geg^{-1}  \\
A &\mapsto & gAg^{-1}+g dg^{-1} \\ 
q &\mapsto & gq g^{-1} \\
\lambda &\mapsto& g\lambda, 
\end{array}
\ee
where $g\in  C^{\infty}(\sM,\SU(2))$ and
\be
\label{gravy2} 
\begin{array}{ccl}
e&\mapsto& e+d_A\eta \\
q &\mapsto&q -\eta, \\
\end{array}
\ee
where $\eta\in C^{\infty}(\sM,\su(2))$. In addition to these gauge
symmetries, the action is invariant under $\lambda\mapsto
\lambda h$ where $h \in C^{\infty}(\gamma,H)$ and $H\subset
\SU(2)$ is the subgroup consisting of elements $g \in \SU(2)$
that stabilize the vector $v$, meaning that $gvg^{-1} = v$.  The action 
is also invariant under reparametrization of the world-line $\gamma$.

A generalization of the naive action (\ref{naive3d}) to arbitrary
dimensions can be constructed as follows. Let $G$ be a Lie group
such that its Lie algebra $\g$ is equipped with an inner product
invariant under the adjoint action of $G$.  Let $\sM$ be a
$d$-dimensional smooth oriented manifold equipped with an oriented
$(d-2)$-dimensional submanifold $\sW$, which we call the `membrane
world-sheet'.  Let $P$ be a principal $G$-bundle over $M$; to
simplify the discussion we shall assume $P$ is trivial, but this
is not essential.  One can define the action 
\be 
S(A,B)=\int_{\sM} \tr[B\wedge F(A)] + {\tau} \int_{\sW} \tr[B\, v] 
\ee 
where $\tau$
is the membrane tension, $B$ is a $\g$-valued $(d-2)$-form, $A$ is
a connection on $P$, $v$ is a fixed but arbitrary unit vector in
$\g$, and `$\tr$' stands for the invariant inner product in $\g$.
The first term is the standard $BF$ theory action, while the
second is a source term that couples $BF$ theory to the membrane
world-sheet.

As with the action in equation (\ref{naive3d}), the above action
is only gauge-invariant if we restrict gauge transformations to be
trivial on the membrane world-sheet.  We can relax this condition
by introducing appropriate degrees of freedom for the
$(d-3)$-brane whose world-sheet is $\sW$. The resulting action is:
\be 
\label{membrane} 
S(A,B,q,\lambda)= \int_{\sM} \tr[B \wedge F(A)] + 
{\tau} \int_{\sW} \tr[(B + d_Aq) \, \lambda  v \lambda^{-1}], 
\ee 
where $q$ is a $\g$-valued $(d-3)$-form on
$\sW$ and $\lambda$ is a $G$-valued function on $\sW$.

This action is invariant under the gauge transformations:
\be
\label{gravy3} 
\begin{array}{ccl}
B &\mapsto & gBg^{-1}  \\
A &\mapsto & gAg^{-1}+g dg^{-1} \\ 
q &\mapsto & gq g^{-1} \\
\lambda &\mapsto& g\lambda, 
\end{array}
\ee
where $g\in  C^{\infty}(\sM,G)$ and
\be
\label{gravy4} 
\begin{array}{ccl}
B&\mapsto& B+d_A\eta \\
q &\mapsto&q -\eta, \\
\end{array}
\ee
where $\eta$ is any $\g$-valued $(d-3)$-form.  As in the
particle case, the action is also invariant under $\lambda\mapsto
\lambda h$, where $h\in C^{\infty}(\sW,H)$ and $H \subseteq
G$ is the subgroup stabilizing $v$, and 
under reparametrization of the membrane world-sheet.

Perhaps the most intuitive equation of motion comes from varying
the $B$ field.  This says that the connection $A$ is flat except
at $\sW$: 
\be 
F = -p  \delta_{\sW} , 
\ee 
where $p = {\tau} \lambda v \lambda^{-1}$ 
and $\delta_{\sW}$ is the distributional
2-form (current) associated to the membrane world-sheet. So, the
membrane causes a conical singularity in the otherwise flat
connection $A$.  The strength of this singularity is determined by
the field $p$, which plays the role of a `momentum density' for
the brane.  Note that while the connection $A$ is singular in the
directions transverse to $\sW$, it is smooth and indeed flat when 
restricted to $\sW$.  Thus the equation of motion obtained from
varying $q$ makes sense:
\be
d_A p = 0 .
\ee
This expresses conservation of momentum density.

\section{The canonical analysis for $d=4$}\label{si}

In this section we work out the other equations of motion
as part of a canonical analysis of the action (\ref{membrane}).  
But, in order to simplify the presentation, we restrict for 
the moment to the case $d=4$---that is, the coupling of a 
string to four-dimensional $BF$ theory.  In Section \ref{gene}, 
we generalize the calculations to arbitrary dimensions.

For this canonical analysis, we assume the spacetime manifold is of the form
${\sM}=\Sigma \times \R$.   We choose local
coordinates $(t,x^a)$ for which $\Sigma$ is given as the
hypersurface $\{t=0\}$.  By definition, $x^a$ with $a=1,2,3$ are local
coordinates on $\Sigma$. We also choose local coordinates
$(t,s)$ on the $2$-dimensional world-sheet $\sW$,
where $s \in [0,2\pi]$ is a coordinate along the one-dimensional
string formed by the intersection of $\sW$ with $\Sigma$.
We pick a basis $e_i$ of the Lie algebra
$\g$, raise and lower Lie algebra indices using the inner product,
and define structure constants by $[e_i, e_j] = c_{ij}^k e_k$.

Performing the standard Legendre transformation one obtains
$E^a_i=\epsilon^{abc} B_{ibc}$ as the momentum canonically
conjugate to $A_a^i$.  Similarly, $\pi^a_i= {\tau} \frac{\partial
x^a}{\partial s} \tr[e_i\lambda v\lambda^{-1}]$ is the momentum
canonically conjugate to $q_a^i$.  This is a version of the $p$
field mentioned in the previous section. There are also certain
fields $\sigma_i$ defined on the string, which are
essentially\footnote{The field $\lambda$ takes values in the group
$G$, so if we think of it as a kind of `position' variable,
position-momentum pairs lie in $T^\ast G$.  Each basis element
$e_i$ of $\g$ gives a left-invariant vector field on $G$ and thus
a function $\sigma_i$ on $T^\ast G$, which describes one component
of the `momentum'. The usual symplectic structure on $T^\ast G$
gives
\[\{\sigma_i,\sigma_j\}=c_{ij}^k \sigma_k ,\]
but recalling that $\lambda$ and thus its conjugate momentum
is actually a function of the coordinate $s$ on the string world-sheet,
we expect
\[\{\sigma_i(s),\sigma_j(s')\}=c_{ij}^k \sigma_k(s) \delta^{(1)}(s-s') \]
and indeed this holds, in analogy to Sousa Gerbert's \cite{1}
calculation for the three-dimensional case.} the momenta conjugate
to $\lambda$. These phase space variables satisfy the following
primary constraints: 
\be 
\label{uno} 
\sigma_i= 0 
\ee 
\be 
\label{tres}
\pi^a_i = {\tau} \frac{\partial x^a}{\partial s}  \,
\tr[e_i\lambda v\lambda^{-1}] 
\ee 
\be 
\label{redu} 
D_a\pi^a_i=0
\ee 
\be 
\label{gauss} 
D_aE^a_i = 
\int_{\Ss}c_{ij}^{k} q^j_a \pi^a_k \ \delta^{\va (3)}(x-x_{\va \Ss}(s)) 
\ee 
\be
\label{curvature} 
\epsilon^{abc}F_{ibc}(x) = 
- \int_{\Ss} \pi^a_i \ \delta^{\va (3)}(x- x_{\va \Ss}(s)). 
\ee 
Here $\Ss\subset\Sigma$
denotes the one-dimensional curve representing the string,
parametrized by $x_{\va \Ss}(s)$.  Equation (\ref{uno}) expresses
the fact that no time derivatives of $\lambda$ appear in the
action.   Equation (\ref{tres}) relates the
conjugate momentum $\pi$ to the field $\lambda$.
The constraint (\ref{redu}) implies that the momentum $\pi_i^a$ is
covariantly constant along the string.  This constraint is
redundant, since it could be obtained by taking the covariant
derivative of (\ref{curvature}) and applying the Bianchi identity.
However, this argument requires some regularization due to the
presence of the $\delta$ distribution on the right.  The
constraint (\ref{gauss}) is the modified Gauss law of $BF$ theory
due to the presence of the string.

Finally, (\ref{curvature}) is the modified curvature constraint
containing the dynamical information of the theory.  This
constraint implies that the connection $A$ is flat away from the
string $\Ss$.    Some care must be taken to correctly intepret the
constraint for points on $\Ss$.  By analogy with the case of 3d
gravity, the correct interpretation is that the holonomy of an
infinitesimal loop circling the string at some point $x \in \Ss$ is
$\exp(-p(x)) \in G$, where $p = \tau \lambda v \lambda^{-1}$ as
before.  This describes the conical singularity of the connection 
at the string world-sheet.

The $BF$ phase space variables satisfy the standard commutation
relations:
\be
\{E_i^a(x),A_b^j(y)\}=\delta_b^a\delta_i^j \,
\delta^{\va(3)}(x-y)
\ee
\be
\{E_i^a(x),E^b_j(y)\}=\{A_a^i(x),A_b^j(y)\}=0.
\ee

Concerning the string canonical variables, there are second class
constraints (this can be seen from the consistency conditions which
say that the time derivatives of (\ref{uno}) and (\ref{tres}) vanish). 

They
can be solved in a way analogous to the point particle case
\cite{1,2}. As in the latter, this leads to a convenient
parametrization of the phase space of the string in terms of the
momentum $\pi^a_i$ and the `total angular momentum' $J_i =
c_{ij}^{k} q^j_a \pi^a_k+\sigma_i$. The Poisson brackets of these
variables are given by 
\be 
\{\pi^a_i(s),J_j(s')\} = 
c_{ij}^{k} \pi^a_k(s) \delta^{\va (1)}(s-s') 
\ee 
\be 
\{J_i(s),J_j(s')\} =
c_{ij}^{k} J_k(s) \delta^{\va (1)}(s-s'). 
\ee 
It is important to
calculate the Poisson bracket\footnote{The presence of second
class constraints in the phase space of the string means that
instead of the standard Poisson bracket one should use the
appropriate Dirac bracket. However, due to the fact that both
$\pi^a_i$ and $J_i$ commute with the constraints, the Dirac
bracket and the standard Poisson bracket coincide for the previous
two equations as well as for the following one.} 
\be 
\label{***}
\{J_i(s),\lambda(s')\} = 
-e_i\lambda(s) \delta^{\va (1)}(s-s'). 
\ee
The string variables are still subject to the following first
class constraints: 
\be 
\label{spin} 
\tr[e_i\lambda z\lambda^{-1}]J^i=0 \ee 
\be
\label{shell} 
\tr[\pi^a\lambda z\lambda^{-1}] = 
\tau \frac{\partial x^a}{\partial s} \tr[v z], 
\ee where $z\in\g$ is such that $[z,v]=0$. 
The last constraint is the
generalization of the mass shell condition for point
particles in 3d gravity.

The Poisson bracket of the string variables with the $BF$ variables
is trivial, as well as the Poisson brackets among the $\pi^a_i$.
In the next section we shall find a representation of the previous
variables as self-adjoint operators acting on an auxiliary Hilbert
space $\Ha$. The constraints above will also be quantized and
imposed on $\Ha$ in order to construct the physical Hilbert space $\Hp$.

\section{Quantization}
\label{quantization}

The auxiliary Hilbert space has the tensor product structure
\[\Ha=\Hb_{\va BF}\otimes \Hb_{\va ST}, \]
where $\Hb_{\va BF}$ and $\Hb_{\va ST}$ are the $BF$ theory
and string auxiliary Hilbert spaces, respectively. In the following
two subsections we describe the construction of such Hilbert
spaces; in the third we define the so-called kinematical Hilbert
space $\Hk$ by quantizing and imposing all the constraints but the
curvature constraint (\ref{curvature}). In the last subsection we
sketch the definition of the physical Hilbert space.

\subsection{The $BF$ auxiliary Hilbert space}

When the group $G$ is compact, we may quantize
the $BF$ theory degrees of freedom just as in standard
loop quantum gravity.  For this reason we only provide a quick
review of how to construct the relevant Hilbert space.  A detailed
description of this construction can be found in \cite{3}.

Briefly, the auxiliary Hilbert space for $BF$ theory, $\Hb_{\va BF}$,
is given by $L^2(\bar{\cal A},\mu)$ where $\bar{\cal A}$ is a certain
completion of the space ${\cal A}$ of smooth connections on $P$,
and $\mu$ is the standard gauge- and diffeomorphism-invariant measure
on $\bar{\cal A}$.  A bit more precisely, the construction goes as follows.

One starts from a certain algebra $\Cyl_{\va BF}$ of
so-called `cylinder functions' of the connection $A$.
The basic building blocks of this algebra are
the holonomies $h_{\gamma}(A)\in G$ of $A$ along paths
$\gamma$ in the manifold $\Sigma$ representing space:
\be
\label{hol}
h_\gamma(A) = P \exp\left( -\int_\gamma A \right)
\ee
where $P$ stands for the path-ordered exponential.
An element of $\Cyl_{\va BF}$ is a function
$$\Psi_{\gamma,f} \colon {\cal A} \to \C, $$
where $\gamma$ is a finite directed graph embedded in $\Sigma$ and
$f \colon G^m \rightarrow\C$ is any continuous function, $m$ being
the number of edges of $\gamma$. This function $\Psi_{\gamma,f}$
is given by \be \label{cyl} \Psi_{\gamma,f}(A) =
f(h_1(A),\dots,h_m(A)) \ee where $h_i(A)$ is the holonomy along
the $i$th edge of the graph $\gamma$ and $m$ is the number of
edges.

Given any larger graph $\gamma'$ formed by adding vertices and edges to
$\gamma$, the function $\Psi_{\gamma,f}$ equals
$\Psi_{\gamma^{\prime},f^{\prime}}$ for some continuous function $f^\prime
\colon G^{m'}\rightarrow\C$, where $m'$ is the number of edges of
$\gamma'$.  Using this fact, we can define an inner product on cylinder
functions.  Given any two elements of $\Cyl_{\va BF}$, we can write them as
$\Psi_{\gamma,f}$ and $\Psi_{\gamma,g}$ where $\gamma$ is a sufficiently
large graph.  Their inner product is then defined by:
\ba \label{innerk}
\left\langle \Psi_{\gamma,f}, \Psi_{\gamma,g} \right\rangle
=  \int_{G^m}
\overline{f(h_1,\dots,h_m)} \, g(h_1,\dots,h_m) \;
dh_1 \cdots dh_m
\ea
where $dh_i$ is the normalized Haar measure on $G$.

The auxiliary Hilbert space $\Hb_{\va BF}$ is defined as the Cauchy
completion of $\Cyl_{\va BF}$ under the inner product in (\ref{innerk}).
Using projective techniques it has been shown \cite{3} that $\Hb_{\va BF}$
is also the space of square-integrable functions on a certain space
$\bar{\cal A}$ containing the space ${\cal A}$ of smooth connections on
$\Sigma$.  Elements of $\bar{\cal A}$ are called `generalized connections'.
The measure $\mu$ in equation (\ref{innerk}) is actually a measure on
$\bar{\cal A}$, and we have $\Hb_{\va BF}= L^2(\bar{\cal A},\mu)$.  In
other words, we have
\be
\left\langle \Psi_{\gamma,f}, \Psi_{\gamma,g} \right\rangle =
\int_{\bar{\cal A}} \overline{\Psi_{\gamma,f}(A)} \,
\Psi_{\gamma,g}(A) \; d\mu(A) .
\ee

The (generalized) connection is quantized by promoting the
holonomy (\ref{hol}) to an operator acting by multiplication on
cylinder functions as follows: \be \widehat{h_\gamma(A)} \Psi(A) \; =
\; h_\gamma(A) \Psi(A)\;. \label{ggcc} \ee It is easy to check
that this defines a self-adjoint operator on $\Hb_{\va BF}$.
Similarly, the conjugate momentum $E^a_j$
is promoted to a self-adjoint operator-valued
distribution that acts by differentiation on smooth cylinder
functions, namely:
\be
\hat{E}^a_j= -i\frac{\delta}{\delta A^j_a}.
\ee

Next, one can introduce an orthonormal basis of states in $\Hb_{\va BF}$
using harmonic analysis on the compact group $G$.  Thanks to the
Peter--Weyl theorem, any continuous function $f\colon G\rightarrow \C$ can
be expanded as follows:
\be \label{PW} f(g)=\sum
\limits_{\rho \in \Irrep(G)}
\left\langle f_\rho, \rho(g)\right\rangle .  \ee
Here $\Irrep(G)$ is a set of unitary irreducible representations of $G$
containing one from each equivalence class.  For any $g \in G$, a
representation $\rho \in \Irrep(G)$ gives a linear transformation $\rho(g)
\colon H_\rho \to H_\rho$ for some finite-dimensional
Hilbert space $H_\rho$.  We may think of $\rho(g)$ as an element of the
Hilbert space $H_\rho \otimes H_\rho^\ast$.  The `Fourier
component' $f_\rho$ is another element of $H \otimes
H^\ast$, and $\langle f_\rho, \rho(g)\rangle$ is their inner product.

The straightforward generalization of this
decomposition to functions $f\colon G^m \rightarrow \C$ allows us
to write any cylindrical function (\ref{cyl}) as:
\be
\Psi_{\gamma,f}(A)=
\sum\limits_{\rho_1, \dots, \rho_m \in \Irrep(G)} \;
\prod_{i=1}^m
\left\langle f_{\rho_i}, \, \rho_i(h_i(A))\right\rangle ,
\label{open_spin_network} \ee
where the `Fourier component' $f_{\rho_i}$ associated
to the $i$th edge of the graph $\gamma$ is an element of
$H_{\rho_i} \otimes H_{\rho_i}^\ast$.  We call the functions
appearing in this sum {\em open spin networks}.  A general
open spin network is of the form
\be
\Psi_{\gamma, \vec{\rho}, \vec{f}}(A) \; = \;
\prod_{i=1}^m \left\langle f_{\rho_i}, \rho_i(h_i(A))\right\rangle  .
\ee
Here $\vec{\rho}$ is an abbreviation for the list of representations
$(\rho_1, \dots, \rho_m)$ labelling the edges of the graph,
and $\vec{f}$ is an abbreviation for the tensor product
$f_{\rho_1} \otimes \cdots \otimes f_{\rho_m}$
Note that $\Psi_{\gamma, \vec\rho, \vec f}$ depends in
a multilinear way on the vectors $f_{\rho_i}$, so it indeed
depends only on their tensor product $\vec f$.

\subsection{The string auxiliary Hilbert space}

The auxiliary Hilbert space for the string degrees of
freedom, $\Hb_{\va ST}$, is obtained in an
analogous fashion.  Just as we built the
auxiliary Hilbert space for $BF$ theory starting from continuous
functions of the connection's holonomies along edges in space, we build
the space $\Hb_{\va ST}$ starting from continuous functions of the
$\lambda$ field's values at points on the string.
This space $\Hb_{\va ST}$ can be described as
$L^2(\bar{\Lambda},\mu_{\va ST})$, where $\bar{\Lambda}$
is a certain completion of the space of $G$-valued functions on the
string $\Ss$, and $\mu_{\va ST}$ is the natural measure on this
space.

To achieve this, we first define an algebra
$\Cyl_{\va ST}$ of `cylinder functions' on the space of
$\lambda$ fields, $\Lambda = C^\infty(\Ss, G)$.
An element of $\Cyl_{\va ST}$ is a function
$$\Phi_{p,f} \colon \Lambda \rightarrow \C , $$
where $p = \{p_1, \dots, p_n\}$ is a finite set of points in $\Ss$
and $f \colon G^{n}\rightarrow\C$ is any continuous function.
This function $\Phi_{p,f}$ is given by
\be \label{cylst}
\Phi_{p,f}(\lambda) = f(\lambda(p_1),\dots,\lambda(p_n)) .
\ee

As in the previous section, if $p'$ is a finite set of points in
$\Ss$ with $p \subset p^{\prime}$, then the function $\Phi_{p,f}$ is equal
to $\Phi_{p^{\prime},f^{\prime}}$ for some continuous function $f^\prime
\colon G^{n'}\rightarrow\C$.  This lets us define an inner product on
$\Cyl_{\va ST}$.  Given any two cylinder functions, we can write them as
$\Phi_{p,f}$ and $\Phi_{p,g}$ where $p$ is a sufficiently large finite set
of points in $\Ss$.  We define their inner product by
\be \label{innerkst}
\left\langle \Phi_{p,f}, \Phi_{p,g} \right\rangle
=  \int_{G^n}
\overline{f(h_1,\dots,h_n)} \,
g(h_1,\dots,h_n) \;
dh_1 \cdots dh_n
\ee
where $dh_i$ is the normalized Haar measure on $G$.  One can check that
this is independent of the choices involved.

The auxiliary Hilbert space $\Hb_{\va ST}$ is then defined to be the Cauchy
completion of $\Cyl_{\va ST}$ under this inner product.  Using
projective techniques \cite{3} it has been shown that $\Hb_{\va ST}$ is
$L^2(\bar{\Lambda},\mu_{\va ST})$ for some measure $\mu_{\va ST}$ on a
certain space $\bar{\Lambda}$ containing the space $\Lambda$:
\be
\left\langle \Phi_{p,f}, \Phi_{p,g} \right\rangle =
\int_{\bar{\Lambda}} \overline{\Phi_{\gamma,f}(\lambda)} \,
\Phi_{\gamma,g}(\lambda) \; d\mu_{\va ST}(\lambda).
\ee
In fact, $\bar{\Lambda}$ is just the space of {\it all} functions
$\lambda \colon \Ss \to G$.  Though very large, this is actually a
compact topological group by Tychonoff's theorem, and $\mu_{\va ST}$
is the Haar measure on this group.

The field $\lambda$ is quantized in terms of operators acting by
multiplication in $\Hb_{\va ST}$. Therefore, the wave functional
$\Phi(\lambda)$ gives the momentum representation of the quantum
state of the string. More precisely, in this representation the
momentum operator $\pi^a_i= {\tau} \frac{\partial x^a}{\partial s}
\tr[e_i\lambda v\lambda^{-1}]$ acts by multiplication, namely: 
\be
\widehat{\pi^a_i(\lambda)} \Phi(\lambda) \; = \; 
{\tau} \frac{\partial x^a}{\partial s}\tr[e_i\lambda v\lambda^{-1}]
\Phi(\lambda). 
\label{ggcc2} 
\ee 
It is easy to check that the
momentum operator is self-adjoint on $\Hb_{\va BF}$. According to
(\ref{***}), the `total angular momentum' $J_i \equiv c_{ij}^{k}
q^j_a \pi^a_k+\sigma_i$ is promoted to a self-adjoint
operator-valued distribution that acts as a derivation, namely 
\be
J^j=-i\frac{\delta}{\delta \lambda^j} . 
\ee 

An application of harmonic analysis on the group $G$, analogous
to what was done in the previous section, lets us write any
cylinder function (\ref{cylst}) as
\be
\Phi_{p,f}(\lambda)=
\sum\limits_{\rho_1, \dots, \rho_n \in \Irrep(G)}  \;
\prod_{i = 1}^n
\left\langle f_{\rho_i}, \rho_i(\lambda(p_i)) \right\rangle ,
\label{point_spin_state}
\ee
where $\rho_i$ runs over irreducible unitary
representations of $G$ on finite-dimensional Hilbert spaces
$H_{\rho_i}$, and the `Fourier component' $f_{\rho_i}$ is an element of
$H_{\rho_i} \otimes H_{\rho_i}^\ast$.  We call the
functions appearing in the sum {\em $n$-point spin states}.
A typical $n$-point spin state is of the form
\be     \Phi_{p, \vec{\rho}, \vec{f}}(\lambda)  \; = \;
\prod_{i=1}^n
\left\langle f_{\rho_i}, \rho_i(\lambda(p_i))\right\rangle .
\ee
Here $\vec{\rho}$ is an abbreviation for the list of representations
$(\rho_1, \dots, \rho_n)$ labelling the points in $p$,
and $\vec{f}$ is an abbreviation for the tensor product
$f_{\rho_1} \otimes \cdots \otimes f_{\rho_n}$.

We hope the strong similarity between the $BF$ and
string auxiliary Hilbert spaces is clear.  The only real
difference is that the $A$ field assigns group elements to
edges, while the $\lambda$ field assigns group elements to
points.  So, we need 1-dimensional spin networks
to describes states of $BF$ theory, but their 0-dimensional
analogues for the $\lambda$ field.

\subsection{The kinematical Hilbert space}

The next step in the Dirac program is to implement the first class
constraints found above as operator equations in order to define
the physical Hilbert space. Here we implement the constraints
(\ref{gauss}), (\ref{spin}), and (\ref{shell}). The states in the
kernel of these quantum constraints define a proper subspace of
$\Ha$ that we call the {\em kinematical Hilbert space}
\[          \Hk \subset\Ha = \Hb_{\va BF} \otimes \Hb_{\va ST} .\]
The implementation of the remaining curvature
constraint (\ref{curvature}) (which also implies (\ref{redu}))
will be discussed in the next subsection.

The constraint (\ref{shell}) is automatically satisfied. This can
be easily checked using the fact that one is working in the
momentum representation where equation (\ref{ggcc2}) holds.

The Gauss constraint (\ref{gauss}) acts on the connection $A$
generating gauge transformations $g \in C^\infty(\Sigma,G)$ whose action
transforms the holonomies along edges of any graph as follows:
\be
\label{gg}
h_e(A) \mapsto g(s(e)) \, h_e(A) \, g(t(e))^{-1}
\ee
where $s(e), t(e) \in \Sigma$ are the {\em source} and {\em target}
vertices of the edge $e$ respectively.
As a result, such gauge transformations act on open spin networks
in $\Hb_{\va BF}$ as follows:
\be
\prod_{i=1}^n
\left\langle f_{\rho_i}, \, \rho_i(h_i(A)\right\rangle
\; \mapsto \;
\prod_{i=1}^n
\left\langle f_{\rho_i}, \, \rho_i(g(s(e_i))h_i(A)g(t(e_i))^{-1})\right\rangle .
\ee
Such gauge transformations also act on the $\lambda$ field:
\be
\lambda \mapsto g\lambda,
\ee
so they act on $n$-point spin states in $\Hb_{\va ST}$ as
follows:
\be
\prod_{i=1}^n
\left\langle f_{\rho_i} , \,\rho_i(\lambda(p_i))\right\rangle
\; \mapsto  \;
\prod_{i=1}^n
\left\langle f_{\rho_i} , \, \rho_i(g(p_i) \lambda(p_i))\right\rangle  .
\ee
Combining these representations, we obtain a
unitary representation of the group $C^\infty(\Sigma,G)$
on $\Ha = \Hb_{\va BF} \otimes \Hb_{\va ST}$.   Gauge-invariant
states are those invariant under this action.

A spanning set of gauge-invariant states can then be constructed in
analogy with the known construction for 3d quantum gravity coupled
to point particles \cite{ky}.  We form such states by taking the
tensor product of an open spin network $\Psi_{\gamma, \vec \rho, \vec f}$
and an $n$-point spin state $\Phi_{p, \vec{\rho'}, \vec{f}'}$.
Such a tensor product state will be invariant under the action
of $C^\infty(\Sigma,G)$ if we:

\begin{enumerate}
\item Require the graph $\gamma$ for the open spin network
to have vertices that include the points $\{p_1, \dots, p_n\}$ forming
the set $p$.

\item Associate an intertwining operator to
each vertex $v$ of the graph $\gamma$ as follows:

a) If the vertex $v$ is not on the string, then choose an
intertwining operator
\[   \iota_{v} \colon \rho_{i_1} \otimes \cdots \otimes \rho_{i_t}
\to \rho_{j_1} \otimes \cdots \otimes \rho_{j_s}  , \]
where $i_1, \dots i_t$ are the edges of $\gamma$ whose target is
$v$, and $j_1, \dots j_s$ are the edges of $\gamma$ whose source
is $v$.

b) If the vertex $v$ is on the string, it coincides with
some point $p_k \in p$.   Then choose an intertwining operator
\[   \iota_{v} \colon \rho_{i_1} \otimes \cdots \otimes \rho_{i_t}
\to (\rho_{j_1} \otimes \cdots \otimes \rho_{j_s}) \otimes \rho'_k, \]
where $\rho'_k$ is the representation labelling the point $p_k$
in the $n$-point spin state $\Phi_{p,\vec{\rho'},\vec{f'}}$.

\item Tensor all the intertwining operators $\iota_v$.   The
result is an element of
\[
\bigotimes_{i=1}^m (H_{\rho_i} \otimes H^\ast_{\rho_i}) \otimes
\bigotimes_{i=1}^n (H_{\rho'_i} \otimes H^\ast_{\rho'_i}) .
\]
Demand that this equals $\vec f \otimes {\vec{f}}'$.  This
fixes our choice of $\vec f$ for the open spin network and
$\vec{f}'$ for the $n$-point spin state.
\end{enumerate}
One can check that such states actually span the space of
states in $\Hb$ that are invariant under gauge transformations
in $C^\infty(\Sigma, G)$.  So, we have solved the Gauss constraint.

Finally, constraint (\ref{spin}) generates gauge transformations
\be \lambda \mapsto \lambda h
\ee
for any $h \in C^\infty(\Ss, H)$, where $H \subseteq G$
is the subgroup stabilizing the 
vector $v$.  These transformations are unitarily
represented on $\Hb_{\va ST}$.  The gauge transformation $h$ acts on
$n$-point spin functions as follows:
\be
\prod_{i=1}^n
\left\langle f_{\rho_i}, \, \rho_i(\lambda(p_i))\right\rangle
\mapsto
\prod_{i=1}^n
\left\langle f_{\rho_i}, \, \rho_i(\lambda(p_i) h(p_i))\right\rangle  .
\ee
We can find $n$-point spin functions
$\Phi_{p,\vec{\rho'}, \vec{f'}}$ that are invariant under these
transformations by choosing the vectors $\vec{f'}$ in such a
way that each vector $f'_{\rho'_j}$
is invariant under the action of the group $H$.

\begin{figure}[h]
\centerline{\hspace{0.5cm} \(
\begin{array}{c}
\includegraphics[height=6cm]{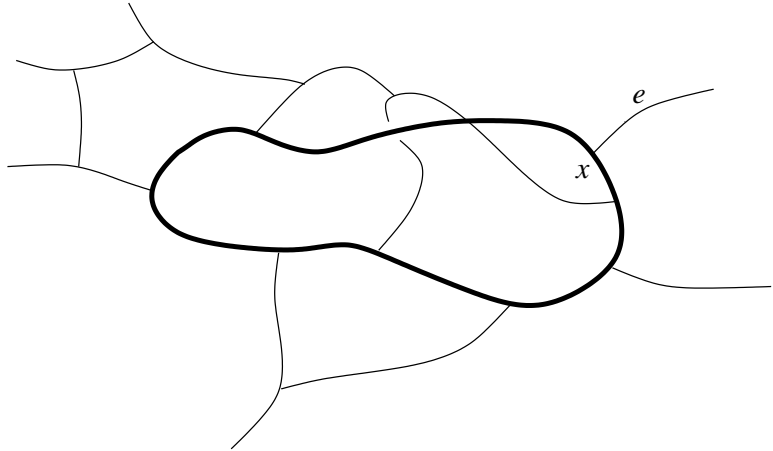}
\end{array}\)} \caption{A typical string spin network.
The Gauss law implies that if a single spin network edge $e$
ends at some point $x$ on the string, the representation
$\rho_e$ is evaluated on the product of the associated holonomy
$h_e(A)$ and the value of the $\lambda$ field at $x$.
}
\label{stringy}
\end{figure}

We call the resulting states
\[           \Psi_{\gamma, \vec\rho, \vec f}
\otimes      \Phi_{p, \vec{\rho'}, \vec{f'}}
\]
{\em string spin networks}.  They span $\Hk$.
A typical string spin network state appears in
Figure \ref{stringy}.  The interplay between
the quantum degrees of freedom in the `bulk' and those on the
string (or membrane, in the general setting of the next section) is
reminiscent of that appearing in the loop quantization of the
degrees of freedom of an isolated horizon in loop quantum gravity
\cite{4}.

\subsection{The physical Hilbert space}

In order to construct the physical Hilbert space $\Hp$ we have to
impose the remaining curvature constraint (\ref{curvature}). This
can be achieved by an application of the techniques developed in
\cite{ky}. The physical inner product can be represented as a sum
over spin foam amplitudes which are a simple generalization of the
amplitudes in three dimensions. The associated state sum
invariants can be directly derived from the canonical perspective
presented here. The details of the construction will be provided
elsewhere.

\section{The general case: membranes coupled to $BF$ theory}
\label{gene}

Let us now describe the phase space of the general case in detail.
Recall that $G$ is a general Lie group with Lie algebra $\g$
equipped with an invariant inner product. Performing the canonical
analysis along the same lines as in Section \ref{si} one obtains
$E^a_i=\epsilon^{a a_1\cdots a_{d-2}} B_{i a_1\cdots a_{d-2}}$
as the momentum canonically conjugate to $A_a^i$, where as before
$i$ labels a basis $e_i$ of $\g$.
The momentum canonically conjugate to $q_a^i$ is given by
\[\pi^{a_1\cdots a_{d-3}}_i= \tau  \,
\frac{\partial x^{[a_1}}{\partial s_1} \frac{\partial
x^{a_2}}{\partial s_2}\cdots \frac{\partial x^{a_{d-3}]}}{\partial
s_{d-3}} \, \tr[e_i\lambda v\lambda^{-1}] ,\]\ where $t,s_1,\dots,
s_{d-3}$ are local coordinates on the membrane world-sheet.  The
Gauss law now becomes:
\be 
\label{gaussg}
D_aE^a_i=\int_{\sB}
c_{ij}^{k} q^j_{a_1\cdots a_{d-3}}\pi^{a_1\cdots a_{d-3}}_k \
\delta^{\va (d-1)}(x-x_{\va \sB}), 
\ee 
where $\sB$ denotes the
brane, i.e.\ the intersection of the membrane world-sheet $\sW$
with $\Sigma$.  The curvature constraint becomes:
\be
\label{curvatureg} 
\epsilon^{a_1\cdots
a_{d-3}bc}F_{ibc}= -\int_{\sB} \pi^{a_1\cdots a_{d-3}}_i \
\delta^{\va (d-1)}(x-x_{\va \sB}) .
\ee 
We also have 
\be
\label{redug} D_a\pi^{a a_1\cdots a_{d-4}}_i=0. 
\ee 
There are
additional constraints for the degrees of freedom of the
$(d-3)$-branes, namely 
\be 
\label{sping} 
\tr[e_i\lambda z\lambda^{-1}]
J^i=0\ \ \ {\rm where}\ \ \ J_i \equiv c_{ij}^{k} q^j_{a_1\cdots
a_{d-3}} \pi^{a_1\cdots a_{d-3}}_k+\sigma_i 
\ee 
and 
\be
\label{shellg} 
\tr[\pi^{a_1\cdots a_{d-3}}\lambda z\lambda^{-1}]
= \tau \frac{\partial x^{[a_1}}{\partial s_1}\frac{\partial
x^{a_2}}{\partial s_2}\cdots \frac{\partial x^{a_{d-3}]}}{\partial
s_{d-3}}\tr[v z], 
\ee
for $[z,v]=0$.

The quantization of the general $d$-dimensional $BF$ theory coupled
to $(d-3)$-branes can be achieved by following an essentially
analogous path as the one described in detail for $4$-dimensional
$BF$ theory coupled to strings.  As long as the gauge group $G$ is compact,
the techniques used in the construction of the auxiliary Hilbert
spaces as well as the definition of the kinematical Hilbert space
and finally the physical Hilbert space can be directly generalized.
In particular, the kinematical Hilbert space is spanned by \emph{membrane
spin networks}, which generalize the string spin networks of the
$4$-dimensional case.

\section{Conclusions}

There are formulations of gravity in four dimensions which are
closely related to $BF$ theory.  The results presented here could
lead to natural candidates for the introduction of matter in those
models. Examples of interest are the MacDowell--Mansouri
formulation of gravity \cite{7}, which is a perturbed version of
$BF$ theory with gauge group $\SO(3,2)$, $\SO(4,1)$ or $\SO(5)$ depending on
the signature of the metric and sign of the cosmological constant.  Another
interesting approach to gravity is the Plebanski formulation,
obtained by imposing extra constraints on $BF$ theory with gauge
group $\SO(3,1)$ or $\SO(4)$.  The well-known Barrett--Crane model
\cite{8} is a tentative quantization of this theory. At least classically,
the $BF$ theories associated to all these theories can be coupled to
strings using the techniques developed here.

When the gauge group $G$ is compact, we can also quantize these theories.
However, for Lorentzian models $G$ is typically not compact.
In the noncompact case it seems there is no good measure on the space
of generalized connections, which precludes the construction of the
auxiliary Hilbert spaces used above. The main obstacle is the
non-normalizability of the Haar measure.  As long as $G$ is
`unimodular'---i.e., as long as it admits a measure invariant
under both right and left translations, as in all the examples
mentioned above---formulas (\ref{innerk}) and (\ref{innerkst})
can still be given a meaning on a fixed graph \cite{5}.   However,
it is no longer possible to promote this inner product to an
inner product on cylindrical functions \cite{6}.  One can still
attempt to deal with the theory in a
more restricted setting by defining it on a fixed cellular decomposition
of spacetime and then showing that physical amplitudes are independent
of this choice.  This is expected for topological theories such as the
ones defined here, but the study of these models still presents interesting
challenges.

Another subtlety of the noncompact case is that while the Lie
algebra $\g$ may still admit an invariant nondegenerate inner product,
this inner product typically fails to be positive definite.   Indeed,
this happens for all noncompact semisimple groups, such as $\SO(p,q)$
for $p + q > 2$.  This affects the interpretation of the action
(\ref{membrane}) for our theory.  Recall that we imposed the normalization
condition $v\cdot v = 1$ for the vector $v \in \g$.   We used this
condition to give a meaning to the tension parameter $\tau$, but
the action only depends on the combination $p = \tau \lambda v \lambda^{-1}$.
As we have seen in the four-dimensional case, the field $p$ has a simple
meaning: the holonomy of the connection $A$ around any small loop encircling
the membrane world-sheet is $\exp(-p) \in G$.  The same is true in any
dimension.

This suggests a simpler action: 
\be
\label{membrane2} 
S(A,B,q,p)=
\int_{\sM} \tr[B\wedge F(A)] +  \int_{\sW} \tr[(B + d_Aq) \, p ],
\ee 
where $p$ is a $\g$-valued function on the world-sheet $\sW$
which under the gauge transformations (\ref{gravy3}) transforms in
the adjoint representation: $p \mapsto g p g^{-1}$. One can check
that the equations of motion still imply $A$ is flat except at
points on $\sW$.  If $\sW$ is connected, this implies that the
holonomy around any small loop encircling the world-sheet is in
the same conjugacy class.  As before, the holonomy around an
infinitesimal loop around some point $x \in \sW$ is 
$\exp(-p(x))$.  It follows that $p$ remains in the same adjoint
orbit over the whole world-sheet.  So, we can write $p$ as $\tau
\lambda v \lambda^{-1}$ for some fixed vector $ v \in \g$ and
some $G$-valued field $\lambda$ on the world-sheet.

When the inner product on $\g$ is positive definite, we can then
fix $\tau$ by normalizing $v$ to have $v \cdot v = 1$.  However,
when the inner product is not positive definite, the new action
(\ref{membrane2}) is more general than the old one, even
for a connected world-sheet, since it allows the momentum density
of the membrane to be space-like ($p\cdot p > 0$) or null
($p\cdot p = 0$), as well as time-like ($p \cdot p < 0$).
One can check that with this new action,
the canonical analysis of Section \ref{si} requires only mild modifications,
and the kinematical construction of the quantum theory presented in
Section \ref{quantization} can still be used, with the precautions
described above for noncompact Lie groups.

It will be interesting to carry out the study of four-dimensional
$BF$ theory coupled to strings in analogy to what has already been
done for three-dimensional gravity coupled to point particles.
For example, point particles in three-dimensional gravity are
known to obey exotic statistics governed by the braid group.
Similarly, we have argued in the companion to this paper that strings
coupled to four-dimensional $BF$ theory obey exotic statistics governed
by the `loop braid group' \cite{baez}.  In that paper we studied these
statistics in detail for the case $G = \SO(3,1)$, but we
treated the strings merely as gauge defects.  It would be good to
study this issue more carefully with the help of the framework
developed here.

\section{Acknowledgements}

A.P.\ would like to thank Karim Noui and Winston Fairbairn for
stimulating discussions.  J.B.\ would like to thank the organizers
of GEOCAL06 for inviting him to Marseille and making
possible the conversations with A.P.\ that led to this paper.


\begin{thebibliography}{10}

\bibitem{spinfoam}
  A.~Perez,
  ``Spin foam models for quantum gravity,''
  Class.\ Quant.\ Grav.\  {\bf 20}, R43 (2003)
  [arXiv:gr-qc/0301113]. D.~Oriti,
  ``Spacetime geometry from algebra: Spin foam models for non-perturbative
  quantum gravity,''
  Rept.\ Prog.\ Phys.\  {\bf 64}, 1489 (2001)
  [arXiv:gr-qc/0106091]. J.~C.~Baez,
  ``An introduction to spin foam models of $BF$ theory and quantum gravity,''
  Lect.\ Notes Phys.\  {\bf 543}, 25 (2000)
  [arXiv:gr-qc/9905087].  J.~C.~Baez,
  ``Spin foam models,''
  Class.\ Quant.\ Grav.\  {\bf 15}, 1827 (1998)
  [arXiv:gr-qc/9709052].


\bibitem{lqg}
T. Thiemann, ``Modern Canonical Quantum General Relativity''
Cambridge, UK: Univ. Pr. (to appear). C. Rovelli,
    `` Quantum gravity,'' Cambridge, UK: Univ. Pr. (2004) 455 p.
A.~Ashtekar and J.~Lewandowski,
  ``Background independent quantum gravity: A status report,''
  Class.\ Quant.\ Grav.\  {\bf 21}, R53 (2004)
  [arXiv:gr-qc/0404018].  A.~Perez,
  ``Introduction to loop quantum gravity and spin foams,''
Proceedings of the International Conference on Fundamental
Interactions, Domingos Martins, Brazil, (2004)
  arXiv:gr-qc/0409061.

\bibitem{ky}
  K.~Noui and A.~Perez,
  ``Three dimensional loop quantum gravity: Coupling to point particles,''
  Class.\ Quant.\ Grav.\  {\bf 22}, 4489 (2005)
  [arXiv:gr-qc/0402111].
  K.~Noui and A.~Perez,
  ``Three dimensional loop quantum gravity: Physical scalar product and  spin
  foam models,''
  Class.\ Quant.\ Grav.\  {\bf 22}, 1739 (2005)
  [arXiv:gr-qc/0402110].  L.~Freidel and D.~Louapre,
  ``Ponzano-Regge model revisited. I: Gauge fixing, observables and
  interacting spinning particles,''
  Class.\ Quant.\ Grav.\  {\bf 21}, 5685 (2004)
  [arXiv:hep-th/0401076].


\bibitem{lau}
L.~Freidel and E.~R.~Livine,
  ``Ponzano-Regge model revisited. III: Feynman diagrams and effective field
  theory,''
  Class.\ Quant.\ Grav.\  {\bf 23}, 2021 (2006)
  [arXiv:hep-th/0502106].
  J.~W.~Barrett,
  ``Feynman diagams coupled to three-dimensional quantum gravity,''
  Class.\ Quant.\ Grav.\  {\bf 23}, 137 (2006)
  [arXiv:gr-qc/0502048].
  J.~W.~Barrett,
  ``Feynman loops and three-dimensional quantum gravity,''
  Mod.\ Phys.\ Lett.\ A {\bf 20}, 1271 (2005)
  [arXiv:gr-qc/0412107].


\bibitem{baez}
  J.~C.~Baez, D.~K.~Wise and A.~S.~Crans,
  ``Exotic statistics for loops in 4d $BF$ theory,''
  arXiv:gr-qc/0603085.

\bibitem{1}
  P.~de Sousa Gerbert,
  ``On Spin And (Quantum) Gravity In (2+1)-Dimensions,''
  Nucl.\ Phys.\ B {\bf 346}, 440 (1990).

\bibitem{2}
  E.~Buffenoir and K.~Noui,
  ``Unfashionable observations about 3 dimensional gravity,''
  arXiv:gr-qc/0305079.

\bibitem{3}
J.\ C.~Baez,
``Generalized measures in gauge theory'', Lett.\ Math.\ Phys.\
{\bf 31} 213, (1994).
  [arXiv:/hep-th/9310201].
J.\ C.~Baez, ``Diffeomorphism-invariant generalized measures
on the space of connections modulo gauge transformations,''
in {\sl Proceedings of the Conference on Quantum Topology},
ed.\ D.\ N. Yetter, World Scientific Press, Singapore, 1994.
  [arXiv:hep-th/9305045].
  A.~Ashtekar and J.~Lewandowski,
  ``Projective techniques and functional integration for gauge theories,''
  J.\ Math.\ Phys.\  {\bf 36}, 2170 (1995)
  [arXiv:gr-qc/9411046]. A.~Ashtekar and J.~Lewandowski, ``Differential geometry on the space of connections via graphs and projective
  limits,'' J.\ Geom.\ Phys.\  {\bf 17}, 191 (1995)
  [arXiv:hep-th/9412073].

\bibitem{4}
  A.~Ashtekar, J.~Baez, A.~Corichi and K.~Krasnov,
  ``Quantum geometry and black hole entropy,''
  Phys.\ Rev.\ Lett.\  {\bf 80}, 904 (1998)
  [arXiv:gr-qc/9710007]. A.~Ashtekar, J.~C.~Baez and K.~Krasnov,
  ``Quantum geometry of isolated horizons and black hole entropy,''
  Adv.\ Theor.\ Math.\ Phys.\  {\bf 4}, 1 (2000)
  [arXiv:gr-qc/0005126].


\bibitem{5}
  L.~Freidel and E.~R.~Livine,
  ``Spin networks for non-compact groups,''
  {\sl J.\ Math.\ Phys.\ } {\bf 44}, 1322 (2003)
  [arXiv:hep-th/0205268].


\bibitem{6}
  J.~L.~Willis,
  ``On the low-energy ramifications and a mathematical extension of loop
  quantum gravity,''
UMI-31-48692.

\bibitem{7} Freidel L. \& Starodubtsev A. (2005).
Quantum gravity in terms of topological observables
[arXiv:hep-th/0501191].

\bibitem{8} Barrett J.W. \& Crane L.  Relativistic spin networks
and quantum gravity, {\it J.\ Math.\ Phys.}, {\bf 39}, 3296 (1998).
[arXiv:gr-qc/9709028].



\end{thebibliography}
\end{document}